 \documentstyle[aps,prl,epsf,amsfonts,amssymb]{revtex}
\begin{document}
\font\Bbb =msbm10  scaled \magstephalf
\def\id{{\hbox{\Bbb I}}}

\title{
Geometry of entangled states }
\author{Marek Ku{\'s}$^{1,3}$ and Karol {\.Z}yczkowski$^{1,2}$}
\address{$^1$Centrum Fizyki Teoretycznej, Polska Akademia Nauk, \\
Al. Lotnik{\'o}w 32/44, 02-668 Warszawa, Poland}
\address{$^2$Instytut Fizyki im. Mariana Smoluchowskiego, \\
Uniwersytet Jagiello{\'n}ski, ul. Reymonta 4, 30-059 Krak{\'o}w, Poland}
\address{$^3$ Laboratoire Kastler-Brossel, Universit\'e Pierre et Marie
Curie, pl. Jussieu 4, 75252 Paris, France}
\date{\today}
\maketitle

\begin{abstract}
Geometric properties of the set of quantum entangled states are
investigated.
We propose an explicit method to compute the dimension of local orbits for
any
mixed state of the general $K\times M$ problem
and characterize the
set of effectively different states (which cannot be
related by local transformations).
Thus we generalize earlier results obtained for
the simplest $2 \times 2$ system,
 which  lead to a stratification of the $6$D set
of $N=4$ pure states.
We define the concept of absolutely separable states,
for which all globally equivalent states are separable.

\end{abstract}

\begin{center}
{\small e-mail: marek@cft.edu.pl \ \quad \  karol@cft.edu.pl}
\end{center}

\section{Introduction}

Recent developments in quantum cryptography and quantum computing evoke
interest in the properties of {\sl quantum entanglement}. Due to recent
works
by Peres \cite{Pe96} and Horodeccy \cite{Ho96} there exist a simple
criterion
allowing one to judge, whether a given density matrix $\rho$, representing a
$2
\times 2 $ or $2 \times 3$ composite system, is separable. On the other
hand,
the general problem of finding sufficient and necessary condition for
separability in higher dimensions remains open 
(see e.g. \cite{LBCKKSST,H300} and references therein).

The question of how many mixed quantum states are separable has been raised
in
\cite{ZHSL,Z99}. In particular, it has been shown that the relative
likelihood
of encountering a separable state decreases with the system size $N$, while
a neighborhood of the maximally mixed state, $\rho_*\sim \id /N$,
remains separable \cite{ZHSL,Z99,BCJLPS}.

From the point of view of a possible applications it is not only important
to
determine, whether a given state is entangled, but also to quantify the
degree
of entanglement. Among several such quantities \cite{LS98,VP98,WT98,Ho99},
the
{\sl entanglement of formation} introduced by Bennet et al. \cite{BVSW96} is
often used for this purpose. Original definition, based on a minimization
procedure, is not convenient for practical use. However, in recent papers of
Hill and Wootters \cite{HW97,Wo98} the entanglement of formation is
explicitly
calculated for an arbitrary density matrix of the size $N=4$.

Any reasonable measure of entanglement have to be invariant with respect to
local transformations \cite{VP98}.  In the problem of $d$ spin $1/2$
particles,
for which $N=2^d$, there exist $4^d-3d+1$ invariants of local
transformations
\cite{LPS99}, and all measures of entanglement can be represented as a
function
of these quantities. In the simplest case $d=2$ there exists $9$ local
invariants, \cite{LPS99,GRB98,EM99,Su00}. These real invariants fix a
state up
to a finite symmetry group and 9 additional discrete invariants (signs) are
needed to make the characterization complete. Makhlin has proved
that two states are locally equivalent if and only if all these $18$
invariants are equal \cite{Ma00}.
 Local symmetry properties of pure states of two and three qubits
where recently analyzed by Carteret and Sudbery \cite{CS00}.
A related geometric analysis of the $2 \times 2$ composed 
system was recently presented by Brody  and Hughston  \cite{BH99}.

The aim of this paper is to  characterize the space of the quantum
"effectively
different" states, i.e. the states non equivalent in the sense of  local
operations. In particular, we are interested in the dimensions and
geometrical
properties of the manifolds of equivalent states.
In a sense our paper is complementary to \cite{CS00}, in which
the authors consider {\sl pure} states for {\sl three} qubits,
while we analyze local properties of {\sl mixed} states
of {\sl two} subsystems of arbitrary size.

We start our analysis
defining in section II the Gram matrix corresponding to any density matrix
$\rho$. We provide an explicit technique of computing the dimension of local
orbits for any mixed state of the general $K\times M$ problem. In section
III
we apply these results to the simplest case of $2 \times 2$ problem. We
describe a stratification of the $6D$ manifold of the pure states and
introduce
the concept of absolute separability. A list of non generic mixed states of
$N=4$ leading to submaximal local orbits is provided in the appendix.

\section{The Gram matrix}

\subsection{$2 \times 2$ system}

For pedagogical reasons we shall start our analysis with the simplest case
of
the $2 \times 2 $ problem. The local transformations of density matrices
form a
six-dimensional subgroup ${\cal L}=SU(2)\otimes SU(2)$ of the full unitary
group $U(4)$. Let $W$ denote a Hermitian density matrix of size $4$,
representing a mixed state. Identification of all states which can be
obtained
from a given one $W$ by a conjugation by a matrix from ${\cal L}$ leads to
the
definition of the "effectively different" states, all effectively equivalent
states being the points on the same orbit of $SU(2)\otimes SU(2)$ through
their
representative $W$.

The manifold $W_{pure}$ of $N=4$ pure states, equivalent to the complex
projective space, ${\mathbb C}P^3$, is $6$ dimensional. Although both the
manifold of pure states and the group of local transformations are
six-dimensional it does not mean that there is only one nontrivial orbit on
${\cal W}_{pure}$. Indeed, at each point $W\in{\cal W}_{pure}$ local
transformations $U({\bf s})$, parametrized by six real variables ${\bf
s}=(s_1,\dots,s_6)$, such that $U({\bf 0})$ equals identity, determine the
tangent space to the orbit, spanned by six vectors:
\begin{equation}
W_i:=\left(\frac{\partial W}{\partial s_i}\right)_{\bf s=0}=
\frac{\partial}{\partial s_i} U({\bf s})WU^\dagger({\bf s})|_{\bf s=0}.
\label{Wi}
\end{equation}
The dimension of the tangent space (equal to the dimension of the orbit)
equals
the number of the independent $W_i$ and, as we shall see, is always smaller
then six. Using the unitarity of $U({\bf s})$ one easily obtains
\begin{equation}
W_i:=\left[\left(\frac{\partial U}{\partial s_i}\right)_{\bf s=0},W\right]
=\left[l_i,W\right], \label{Wi2}
\end{equation}
with $l_i:=\left(\partial U/\partial s_i \right)_{\bf s=0}$, and establishes
the hermiticity of each $W_i$.

Although the so obtained $W_i$ depend on a particular parametrization of
$U({\bf s})$, the linear space spanned by them does not. In fact we can
choose
some standard coordinates in the vicinity of identity for each $SU(2)$
component obtaining
\begin{equation}
l_k=i\sigma_k\otimes I,\quad l_{k+3}=I\otimes i\sigma_k, \label{lk}
\end{equation}
where $\sigma_k$, $k=1,2,3$ stand for the Pauli matrices, and $I$ is the
$2\times 2$ identity matrix. Obviously, the antihermitian matrices $l_i$,
$i=1,\ldots,6$, form a basis of the $\mathfrak{su}_2\oplus\mathfrak{su}_2$
Lie algebra.

The dimensionality of the tangent space can be probed by the rank of the
real
symmetric $6\times 6$ Gram matrix
\begin{equation}
C_{mn}:=\frac{1}{2}\text{Tr}W_mW_n. \label{C}
\end{equation}
formed from the Hilbert-Schmidt scalar products of $W_i$'s in the space of
Hermitian matrices. The most important part of our reasoning is based on
transformation properties of the matrix $C$ along the orbit. In order to
investigate them let us assume thus, that $W'$ and $W$ are equivalent
density
matrices, i.e.\ there exists a local operation $U\in SU(2)\otimes SU(2)$
such
that $W'=UWU^\dagger$. A straightforward calculation shows that the
corresponding matrix $C'$ calculated at the point $W'$ is given by:
\begin{equation}
{C'}_{mn}=\frac{1}{2}\text{Tr}{W'}_m{W'}_n=\frac{1}{2}
\text{Tr}\left(\left[{l'}_m,W\right]\left[{l'}_n,W\right]\right), \label{C1}
\end{equation}
where
\begin{equation}
{l'}_i:=U^\dagger l_i U, \quad i=1,\ldots,6. \label{li1}
\end{equation}

The transformation (\ref{li1}) defines a linear change of basis in the Lie
algebra $\mathfrak{su}_2\oplus\mathfrak{su}_2$ and as such is given by a
$6\times 6$ matrix $O$ i.e.\ ${l'}_i=\sum_{j=1}^6 O_{ij}l_j$. It can be
established that $O$ is a real orthogonal matrix: $O^{-1}=O^T$, either by
the
direct calculation using some parametrization of $SU(2)\otimes SU(2)$
respecting (\ref{lk}), or by invoking the fact that
$\mathfrak{su}_2\oplus\mathfrak{su}_2$ is a real Lie algebra and (\ref{li1})
defines the adjoint representation of $SU(2)\otimes SU(2)$.

Using the above we easily infer that matrices $C$ corresponding to
equivalent
states are connected by orthogonal transformation: $C'=OCO^T$. It is thus
obvious that properties of states which are not changed under local
transformations are encoded in the invariants of $C$, which can thus suit as
measures of the local properties such as entanglement or distilability. As
shown in the following section, the above conclusions remains valid, {\it
mutatis mutandis}, if we drop the condition of the purity of states and go
to
higher dimensions of the subsystems.

\subsection{General case: $ K \times M$ system}

A density matrix $W$ (and, {\it a fortiori}, the corresponding matrix $C$)
of a
general bipartite $K\times M$ system can be conveniently parametrized in
terms
of $(KM)^2-1$ real numbers $a_j$, $b_\alpha$, $G_{j\alpha}$,
$j=1,\ldots,K^2-1$, $\alpha=1,\ldots,M^2-1$ as
\begin{equation}
W=\frac{1}{(KM)^2}I+i a_k(e_k\otimes I)+i b_\alpha(I\otimes f_\alpha)
+G_{k\alpha} (e_k\otimes f_\alpha), \label{WKM}
\end{equation}
where $e_k$ and $f_\alpha$ are generators of the Lie algebras
${\mathfrak{su}}_K$, and ${\mathfrak{su}}_M$ fulfilling the commutation
relations
\begin{equation}
[e_j,e_k]=c_{jkl}e_l, \quad
[f_\alpha,f_\beta]=d_{\alpha\beta\gamma}f_\gamma,
\label{lacomm}
\end{equation}
normalized according to:
\begin{equation}
\text{Tr}e_je_k=-2\delta_{jk},\quad \text{Tr}f_\alpha
f_\beta=-2\delta_{\alpha\beta}. \label{ngen}
\end{equation}
In the above formulas we employed the summation convention concerning
repeated
Latin and Greek indices. We also used the same symbol $I$ for the identity
operators in different spaces, as their dimensionality can be read from the
formulas without  ambiguity. Positivity of the matrix $W$ imposes certain
constraints on on the parameters $a_j$, $b_\alpha$, $G_{j\alpha}$.

By analyzing the effect of a local transformation $L=V\otimes U\in
SU(K)\otimes
SU(M)$ upon $W$ we see that ${\bf a}:=(a_j)$, $j=1,\ldots,K^2-1$ and ${\bf
b}=(b_\alpha)$, $\alpha=1,\ldots,M^2-1$ transform as vectors with respect to
the adjoint representations of $SU(K)$ and $SU(M)$, respectively, whereas
$G:=(G_{i,\alpha})$ is a vector with respect to both adjoint
representations.

In analogy with the previously considered case of pure $2\times 2$ states,
we can choose the parametrization of the local transformations in such a way
that the tangent space to the orbit at $W$ is spanned by the vectors
\begin{eqnarray}
W_i&=&[e_i\otimes I, W], \quad W_\alpha=[I\otimes f_\alpha, W].
\end{eqnarray}
The number of linearly independent vectors equals the dimensionality of the
orbit.
As previously this number is independent of the chosen parametrization and
can be recovered as the rank of the corresponding Gram matrix $C$, which
takes now a block form respecting the division into Latin and Greek indices
\begin{equation}
C=\left[
\begin{array}{cc}
A & B \\ B^T & D
\end{array}
\right], \label{CKM}
\end{equation}
where
\begin{eqnarray}
A_{ij}&=&\frac{1}{2}\text{Tr}W_iW_j,\;
B_{i\alpha}=\frac{1}{2}\text{Tr}W_iW_\alpha,\;
D_{\alpha\beta}=\frac{1}{2}\text{Tr}W_\alpha W_\beta.
\label{blocks}
\end{eqnarray}

The Gram matrix $C$ has dimension $K^2+M^2-2$, the square matrices $A$ and
$D$
are $(K^2-1)$ and $(M^2-1)$ dimensional, respectively, while the rectangular
matrix $B$ has size $(K^2-1) \times (M^2-1)$. The matrix $C$ is nonnegative
definite and the number of its positive eigenvalues gives the dimension of
the
orbit starting at $W$ and generated by local transformations. A direct
algebraic calculation gives
\begin{eqnarray}
A_{ij}&=&(2G_{k\alpha}G_{m\alpha}+Ma_ka_m)c_{ikl}c_{jml}, \nonumber \\
B_{i\alpha}&=&2G_{k\beta}G_{m\gamma}c_{ikm}d_{\alpha\gamma\beta}, \nonumber
\\
D_{\alpha\beta}&=&(2G_{m\gamma}G_{m\delta}+Kb_\gamma
b_\delta)d_{\alpha\gamma\mu}d_{\beta\delta\mu}.
\label{ABDKM}
\end{eqnarray}

In this way we arrived at the main result of this paper:

{\sl
Dimension $D_l$ of the orbit generated by local operations acting on
a given mixed state $W$ of any $K \times M$  bipartite system is equal to
the rank of the Gram matrix $C$ given by (\ref{CKM}) - (\ref{ABDKM}).
}

\medskip

If all eigenvalues of $C$ are strictly positive the local orbit has the
maximal
dimension equal to $D_l=K^2+M^2-2$. In the low dimensional cases it was
always
possible to find such parameters $a_j$ and $b_\alpha$, i.e. such a density
matrix $W$ that the local orbit through $W$ was indeed of the maximal
dimensionality. We do not know if such an orbit exists in an arbitrary
dimension $K\times M$, although we suspect that is the case in a generic
situation (i.e.\ all eigenvalues of $W$ different, nontrivial form of the
matrix
$G$). In the simplest case $2\times 2$ we provide in Appendix A the list of
all, non generic density matrices corresponding to sub-maximal local orbits.
All other density matrices lead thus to the full (six) dimensional local
orbits.

This approach is very general and might be applied for multipartite systems
of any dimension. Postponing these exciting investigations to a subsequent
publication \cite{zasius}, we now come  back to the technically most simple
case of original $2\times 2$-dimensional bipartite system.

\section{Local orbits for the $2 \times 2 $ system}

\subsection{Stratification of the $6D$ space of pure states}

The pure states of a composite $2\times 2$ quantum system form a
six-dimensional submanifold ${\cal W}_{pure}$ of the fifteen-dimensional
manifold of all density matrices in the four-dimensional Hilbert space,
i.e.\
the set of all Hermitian, non-negative $4\times 4$ matrices with the trace
1.
Indeed, the density matrices $W$ and $W'$ of two pure states described by
four-component complex, normalized vectors $|w\rangle\langle w|$ and
$|w'\rangle\langle w'|$ coincide, provided that $|w'\rangle=U|w\rangle$,
where
$U$ is a unitary $4\times 4$ matrix which commutes with $W$. Since $W$ has
threefold degenerate eigenvalue 0, the set of unitary matrices rendering the
same density matrix {\it via} the conjugation $W'=UWU^\dagger$, can be
identified as the six dimensional quotient space $U(4)/[U(3)\times U(1)]
={\mathbb C}P^3$. The manifold of the pure states itself is thus given as
the
set of all matrices obtained from $W_0:=|w_0\rangle\langle w_0|$, where
$|w_0\rangle=[1,0,0,0]^T$, by the conjugation by an element of ${\mathbb
C}P^3$
and conveniently parametrized by three complex numbers $x,y,z$:
$|w\rangle:={\cal N}[1,x,y,z]^T$, $W=W(x,y,z):=|w\rangle\langle w|$, where
${\cal N}=(1+|x|^2+|y|^2+|z|^2)^{-1/2}$ is the normalization constant, and
we
allow the parameters to take also infinite values of (at most) two of them.
In
more technical terms we consider thus the orbit of $U(4)$ through the point
$W_0$ in the space of Hermitian matrices.

In fact, since the normalization of density matrices does not play a role in
the following considerations, we shall take care of it at the very end, and
parametrize the manifold of pure states by four complex numbers $v,x,y,z$
being
the components of $|w\rangle$, (the overbar denotes the complex
conjugation):
\begin{equation}
|w\rangle=\left[
\begin{array}{c}
v \\ x \\ y \\ z
\end{array}
\right], \quad W=|w\rangle\langle w|= \left[
\begin{array}{cccc}
v\bar{v} & v\bar{x} & v\bar{y} & v\bar{z} \\ x\bar{v} & x\bar{x} & x\bar{y}
&
x\bar{z} \\ y\bar{v} & y\bar{x} & y\bar{y} & y\bar{z} \\ z\bar{v} & z\bar{x}
&
z\bar{y} & z\bar{z}
\end{array}
\right], \label{W}
\end{equation}
bearing in mind, when needed, that the sum of their absolute values equals
one.
In fact, equating one of the four coordinates with a real constant yields
one
of four complex analytic maps which together cover the complex projective
space
${\mathbb{C}P}^{3}$ (with which the manifold of the pure states can be
identified) {\it via} standard homogeneous coordinates. This leads to a more
flexible, symmetric notation, and dispose off the need for infinite values
of
parameters.

The dimensionality of the orbit given by $rank(C)$ is the most obvious
geometric invariant of the orthogonal transformations of $C$. As it should
it
does not change along the orbit. All invariant functions (or separability
measures) can be obtain in terms of the functionally independent invariants
of
of the real symmetric matrix $C$ under the action of the adjoint
representation
of $SU(2)\otimes SU(2)$. In particular, the eigenvalues of $C$ are,
obviously,
such invariants. Substituting our parametrization of pure states density
matrices (\ref{W}) to the definition of $C$ (\ref{C}) yields, after some
straightforward algebra, the eigenvalues
\begin{equation}
\lambda_1=0,\quad \lambda_2=8|\omega|^2, \quad
\lambda_3=\lambda_4=1+2|\omega|,\quad \lambda_5=\lambda_6=1-2|\omega|,
\label{spectr}
\end{equation}
where $\omega:=vz-xy$. For any pure state one may explicitly calculate the
entropy of entanglement \cite{BVSW96} or a related quantity, called
concurrence
\cite{Wo98}. For the pure state (\ref{W}) the concurrence equals
\begin{equation}
c=2|\omega|=2|vz-xy| \label{conc}
\end{equation}
and $c\in [0,1]$. Thus the spectrum of the Gram matrix may be rewritten as
\begin{equation}
{\rm eig}(C)=\{0, 2c^2, 1+c,1+c,1-c,1-c \}. \label{spectr2}
\end{equation}
The number of positive eigenvalues of $C$ determines the dimension of the
orbit
generated by local transformation. As already advertised, the dimensionality
of
the orbit is always smaller then  $6$. In a generic case it equals $5$,  but
for $\omega=0$ ($c=0$ - separable states) it shrinks to $4$ and for
$|\omega|=1/2$ ($c=1$ - maximally entangled states) it shrinks to $3$.
These results have already been obtained in a recent
paper by Carteret and Sudbery \cite{CS00},
who have shown that the exceptional states (with local orbits
of a  non-generic dimension)
are characterized by maximal (or minimal) degree of entanglement.

In order to investigate more closely the geometry of
 various orbits let us introduce the
following definition:
\begin{eqnarray}
&{\cal W}_\Omega:=&\{W={\bf w}{\bf w}^\dagger:\, {\bf
w}=[v,x,y,z]^T\in{\mathbb
C}^4, \, \|{\bf w}\|^2=|v|^2+|x|^2 \nonumber \\ & & +|y|^2+|z|^2=1, \;
|(vz-xy)|=\Omega\}. \label{WOmega}
\end{eqnarray}
It is also convenient to define a map from the space of state vectors
$\{{\bf w}=[v,x,y,z]^T\in{\mathbb C}^4 :
\|{\bf w}\|^2=|v|^2+|x|^2+|y|^2+|z|^2=1 \}$ to the space of complex
$2\times 2$ matrices
\begin{equation}
X({\bf w})=\left[
\begin{array}{cc}
v & y \\ x & z
\end{array}
\right]. \label{Xw}
\end{equation}
In terms of $X({\bf w})$ the length of a vector ${\bf w}$ and the bilinear
form
$\omega({\bf w}):=vz-xy$ read thus: $\|{\bf w}\|^2=\text{Tr}X({\bf
w})X^\dagger({\bf w})$ and $\omega({\bf w}) =\text{det}X({\bf w})$. From the
Hadmard inequality
\begin{equation}
|\text{det}X({\bf w})|\leq[(|v|^2+|x|^2)(|y|^2+|z|^2)]^{1/2}, \label{had}
\end{equation}
we infer $|\omega({\bf w})|\leq\frac{1}{2}$. Indeed, since
$|v|^2+|x|^2+|y|^2+|z|^2=1$, the right-hand-side of (\ref{had}) equals its
maximal value of $\frac{1}{4}$ for $|v|^2+|x|^2=\frac{1}{2}=|y|^2+|z|^2$. A
straightforward calculation shows also, that a local transformation
$L=V\otimes
U$ sends ${\bf w}$ to ${\bf w}'=L{\bf w}$ if and only if $X({\bf
w}')=UX({\bf
w})V^T$. As an immediate consequence we obtain the conservation of
$|\omega({\bf w})|$ under local transformation. Together with the obvious
conservation of $\|{\bf w}\|$ (which, by the way, is also easily recovered
from
$\|{\bf w}\|^2=\text{Tr}X({\bf w})X^\dagger({\bf w})$), it shows that the
parametrization (\ref{WOmega}) is properly chosen. Moreover it can be proved
that ${\cal L}$ acts transitively on submanifolds (\ref{WOmega}) of constant
$|\omega|$, i.e.  for each pair $W={\bf w}{\bf w}^\dagger$, $W={\bf w}'{\bf
w}'^\dagger$ such that $|\omega({\bf w})|=|\omega({\bf w}')|=\Omega$, there
exists such a local transformation $L\in{\cal L}$ that
$W'=L(W):=LWL^\dagger$,
or, in other words, that the manifold (\ref{WOmega}) of constant $|\omega|$
is
an orbit of the group of local transformations ${\cal L}$ through a single
point $\bar{W}$ i.e.\ $W_\Omega={\cal L}(\bar{W})$. To this end, it is
enough
to show that each $W\in{\cal W}_\Omega$ can be transformed by a local
transformation into $W_\theta={\bf w}_\theta{\bf w}^\dagger_\theta$, where
${\bf w}_\theta=[\cos(\theta/2),0,0,\sin(\theta/2)]^T$ with $\sin
\theta=2\omega$ (from the above mentioned bound for $|\omega({\bf w})|$ we
know
that it is sufficient to consider $0\leq\theta\leq\frac{\pi}{2}$). To this
end
we invoke the singular value decomposition theorem which states that for an
arbitrary (in our case $2\times 2$) matrix $X$, there exist unitary $U',V'$
such that
\begin{equation}
X':=U'X{V'}^T=\left[
\begin{array}{cc}
p & 0 \\ 0 & q
\end{array}
\right],\quad p\geq q\geq 0. \label{svd}
\end{equation}
Let now $V'=e^{i\xi}V$, $U'=e^{i\eta}U$, $V,U\in SU(2)$. We can rewrite
(\ref{svd}) as
\begin{equation}
UXV^T=\left[
\begin{array}{cc}
pe^{i\phi} & 0 \\ 0 & qe^{i\phi}
\end{array}
\right],\; p\geq q\geq 0 \; \phi:=-(\eta+\xi) \label{svd1}
\end{equation}
Substituting $X=X({\bf w})$ (\ref{Xw}), we obtain
$p^2+q^2=\text{Tr}XX^{\dagger}=\|{\bf w}\|^2=1$ and invoking the
invariance of $pq=|\text{det}X|=|\omega({\bf w})|=\sin 2\theta$. This gives
an
unique solution $p=\cos\theta$, $q=\sin\theta$ in the interval
$0\leq\theta\leq\frac{\pi}{2}$. On the other hand, as above mentioned, the
transformation (\ref{svd1}) corresponds to $L{\bf
w}=w'_\theta=[\cos(\theta/2)e^{i\phi},0,0,\sin(\theta/2)e^{i\phi}]^T$, but
obviously $W'_\theta={\bf w}'_\theta{\bf w}'^\dagger_\theta=W_\theta={\bf
w}_\theta{\bf w}^{\dagger}_{\theta}$, i.e., finally,
$LWL^{\dagger}=W_\theta$,
with $L=V\otimes U\in{\cal L}$ as claimed. This is, obviously, a restatement
of the Schmidt decomposition theorem for $2\times 2$ systems.

Now we can give the full description of the geometry of the states. The line
into $W_\theta={\bf w}_\theta{\bf w}^\dagger_\theta$,
$0\leq\theta\leq\frac{\pi}{2}$ connects all "essentially different" states.
At
each $\theta$ different from $0,\pi/2$ it crosses a five-dimensional
manifold
of the states equivalent under local transformations. The orbits of
submaximal
dimensionality correspond to both ends of the line. For $\theta=\pi/2$ the
orbit is three-dimensional. The states belonging to these orbits are
maximally
entangled, since $|\omega|=1/2$ corresponds to $c=1$.

In order to recover the whole orbits we should find the actions of all
elements
of the group of local transformations on a representative of each orbit
(e.g.
one on the above described line). Since, however, the orbits have dimensions
always lower than the dimensionality of the group, the action is not
effective,
i.e. for each point on the orbit, there is a subgroup of ${\cal L}$ which
leaves this point unmoved. This stability subgroups are easy to identify in
each case. Taking this into account we end up with the following
parametrization of three-dimensional orbits of the maximally entangled
states
\begin{equation}
{\cal W}_{\pi/4}=\{W={\bf w}{\bf w}^{\dagger}:{\bf w}= {\bf
w}(\alpha,\chi_1,\chi_2)\},\quad {\bf
w}(\alpha,\chi_1,\chi_2)\}=\frac{1}{\sqrt{2}}\left[
\begin{array}{l}
\cos\alpha e^{i\chi_1} \\ \sin\alpha e^{i\chi_2} \\ \sin\alpha e^{-i\chi_2}
\\
-\cos\alpha e^{-i\chi_1}
\end{array}
\right], \label{3dtrigm}
\end{equation}
with $0\leq\chi_i<2\pi$, $0\leq\alpha\leq\pi/2$, which means that
topologically
this manifold is a real projective space, ${\mathbb R}P^3=S^3/Z_2$,
where $Z_2$ is a two elements discrete group.
 This is related to
the well known result
that for bipartite systems the maximally entangled states may be produced by
an
appropriate operation performed locally, on one subsystem only. The manifold
of
maximally entangled states (\ref{3dtrigm}) is cut by the line of essentially
different states at the origin of the coordinate system
$(\alpha,\chi_1,\chi_2)$.

The four dimensional orbit corresponding to $\theta=0$ consists of separable
states characterized by the vanishing concurrence,  $c=0$. The
parametrization
of the whole orbit, exhibiting its $S^2\times S^2$ structure, is given by:
\begin{eqnarray}
{\bf w}(\alpha,\beta,\chi_1,\chi_2)= \left[
\begin{array}{l}
\cos\alpha\cos\beta e^{i\chi_1} \\ \cos\alpha\sin\beta e^{i\chi_2} \\
\sin\alpha\cos\beta e^{-i\chi_2} \\ \sin\alpha\sin\beta e^{-i\chi_1}
\end{array}
\right], \nonumber
\\
0\leq\chi_i<2\pi,\; 0\leq\alpha,\beta<\pi/2. \label{4dtrig}
\end{eqnarray}

The majority of states, namely these which are neither separable, nor
maximally
entangled, belongs to various five-dimensional orbits labeled by the values
of
the parameter $\theta$ with $0<\theta<\pi/2$. In this way we have performed
a
stratification of the $6$D manifold of the pure states, depicted
schematically
in Fig.1b.

For comparison we show in Fig. 1a the stratification of a sphere $S^2$,
which
consists of a family of $1$D parallels and two poles. Zero dimensional north
pole on  ${\mathbb C}P^1$ corresponds to the $3$D manifold of maximally
entangled states in ${\mathbb C}P^3$, while the $4$D space of separable
states
may be associated with the opposite pole. In the case of the sphere (the
earth)
the symmetry is broken by distinguishing the rotation axis pointing both
poles.
In the case of $N=4$ pure states the symmetry is broken by distinguishing
the
two subsystems, which determines both manifolds of maximally entangled and
separable states.

\begin{figure}
  \hspace*{3.3cm}
 \epsfxsize=8.0cm
\epsfbox{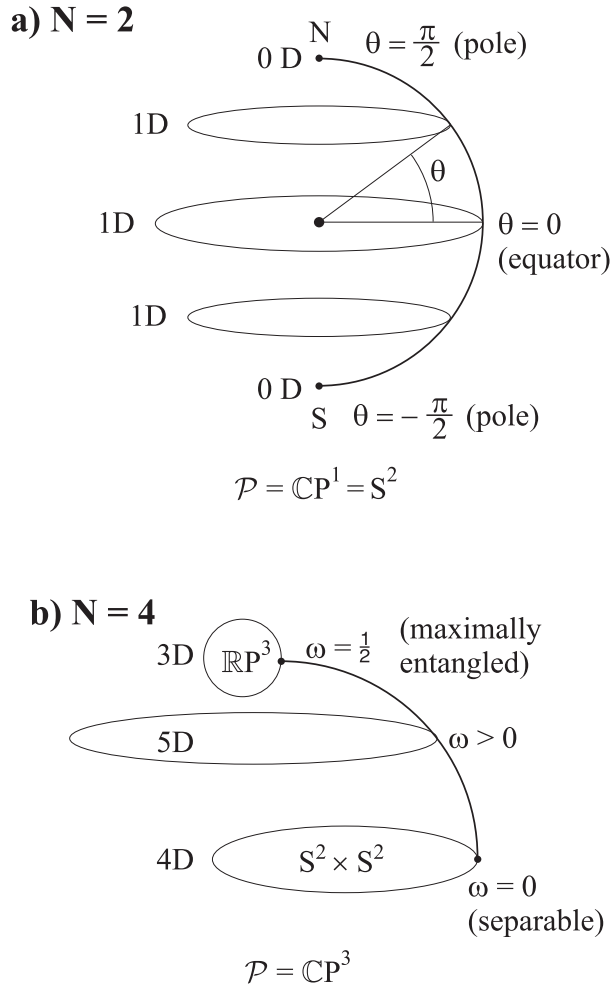} \\ \caption{Stratification of the sphere along the
Greenwich meridian (a), stratification of the $6$ dimensional space of the
$N=4$ pure states along the line of effectively different states, $\omega\in
[0,1/2]$ (b). The poles correspond to the distinguished submanifolds of
${\mathbb C}P^3$: the $3D$ manifold of maximally entangled states and the
$4$D
manifold of separable states. } \label{fig1}
\end{figure}

\subsection{Dimensionality of global orbits}

Before we use the above results to analyze the dimensions of local orbits
for
the mixed states of the $2 \times 2$ problem, let us make some remarks on
the
dimensionality of the global orbits. The action of the entire unitary group
$U(4)$ depends on the degeneracy of the spectrum of a mixed state $W$. Let
$W=VRV^{\dagger }$, where $V$ is unitary and the diagonal matrix $R$
contains
non negative eigenvalues $r_i$.

Due to the normalization condition Tr$W=1$ the eigenvalues satisfy
$r_1+r_2+r_3+r_4=1$. The space of all possible spectra forms thus a regular
tetrahedron, depicted in Fig.2. Without loss of generality we may assume
that $
r_1 \ge r_2 \ge r_3\ge r_4 \ge 0$. This corresponds to dividing the $3$D
simplex into $24$ equal asymmetric parts and to picking one of them. This
set,
sometimes called the {Weyl chamber} \cite{Weylc}, enables us to parametrize
entire space of mixed quantum states by global orbits generated by each of
its
points.

\begin{figure}
  \hspace*{4.3cm}
 \epsfxsize=8.0cm
\epsfbox{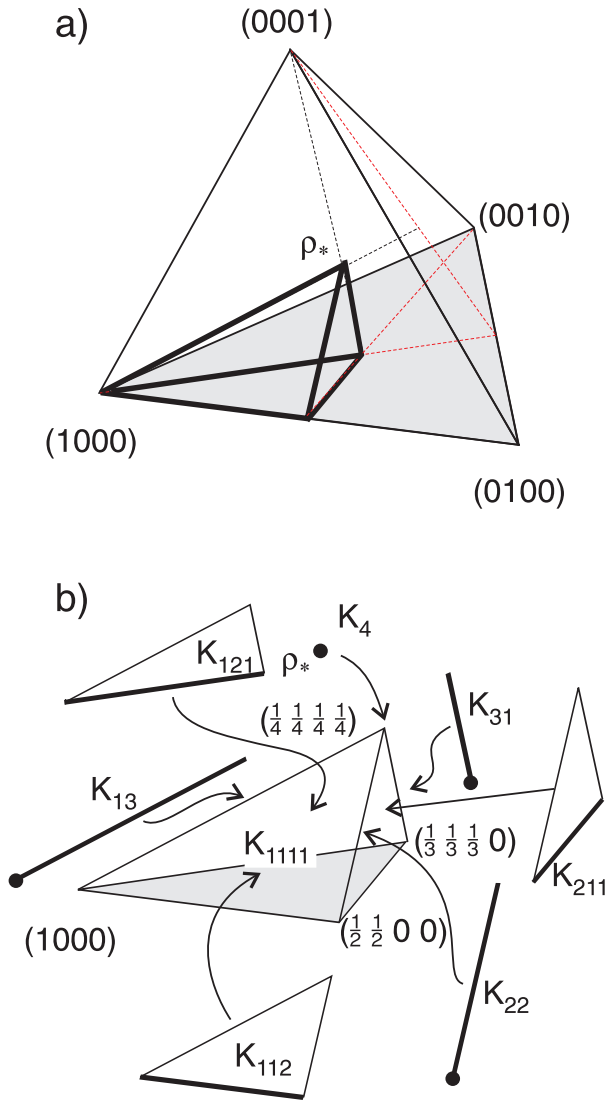} \\ \caption{ The simplex of eigenvalues of the $N=4$
density matrices (a). Pure states are represented by four corners of the
thetrahedron, while its center denotes the maximally mixed state $\rho_*$.
Magnification of the asymmetric part od the simplex,
related to the Weyl chamber (b). It can be decomposed into $8$ parts according
to different kinds of degeneracy of the spectrum. } \label{fig3}
\end{figure}

Note that the unitary matrix of eigenvectors $V$ is not determined uniquely,
since $W =VRV^{\dagger }=VHRH^{\dagger }V^{\dagger }$, where $H$ is an
arbitrary diagonal unitary matrix. This stability group of $U$ is
parametrized
by $N=4$ independent phases. Thus for a generic case of all eigenvalues
$r_i$
different, (which corresponds to the interior $K_{1111}$ of the simplex),
the
space of global orbits has a structure of the quotient group
$U(4)/[U(1)^4]$.
It has $D_g=16-4=12$ dimensions.

If degeneracy in the spectrum of $W$ occurs, say $r_1 =r_2 > r_3 > r_4$,
than
the stability group  $H=U(2)\times U(1)\times U(1)$ is $4+1+1=6$ dimensional
\cite{ACH93}. In this case, corresponding to the face $K_{211}$ of the
simplex,
the global orbit $U/H$ has $D_g=16-6=10$ dimensions. The dimensionality is
the
same for the other faces of the simplex, $K_{121}$ and $K_{112}$. The
important
case of pure states corresponds to the triple degeneracy, $r_1> r_2 = r_3 =
r_4$ for which the stability group  $H$ equals $U(3)\times U(1)$. The orbits
$U/H=SU(4)/U(3)$ have a structure of complex projective space ${\mathbb
C}P^3$.
This $6$D manifold results thus of all points of the Weyl chamber located at
the edge $K_{13}$. These parts of the asymmetric simplex are shown in Fig.2,
the indices labeling each part give the number of degenerated eigenvalues in
decreasing order. For another edge $K_{22}$ of the simplex $H=U(2)\times
U(2)$
and the quotient group $U/H$ is $16-8=8$ dimensional. In the last case of
quadruple degeneracy, corresponding to the maximally mixed state
$\rho_*=\id /4$, the stability group $H=U(4)$, thus $D_l=0$. A
detailed
description of the decomposition of the Weyl chamber with respect to the
dimensionality of global orbits for arbitrary dimensions is provided in
\cite{ZS00}.

\subsection{Dimensionality of local orbits}

For $K=M=2$ (two qubit system) $c_{ijk}=-2\epsilon_{ijk}$ and
$d_{\alpha\beta\gamma}=-2\epsilon_{\alpha\beta\gamma}$, where
$\epsilon_{\alpha
\beta \gamma}$ is completely antisymetric tensor. Formulae (\ref{ABDKM})
give
in this case
\begin{eqnarray}
A=8[(\text{Tr}G'G'^T)\cdot I - G'G'^T]+ 8(\|\bf{a'}\|^2\cdot I-{\bf a'}{\bf
a'}^T) ,\\ D=8[(\text{Tr}G'G'^T)\cdot I - G'^T G']+ 8(\|\bf{b'}\|^2\cdot
I-{\bf
b'}{\bf b'}^T) , \label{AD}
\end{eqnarray}
and
\begin{equation}
BG'^T=G'^TB=-16\;\text{det}G'\cdot I, \label{B}
\end{equation}
where $3D$ vectors $\bf a'$, $\bf b'$  and a $3 \times 3$ matrix  $G'$
represent a certain $N=4$ mixed state $W$ in the form (\ref{WKM}). For later
convenience we denote the system variables by symbols with primes. For
$\text{det}G'\ne 0$ the last equation gives $B=-16\text{det}G'^T\cdot
(G'^T)^{-1}$, but below we will show the more convenient representation of
$B$.

Since $G'$ is real, we can find its singular value decomposition in terms of
two real orthogonal matrices $O_1$, $O_2$ and a positive diagonal matrix
\begin{equation}
O_1G'O_2^T=G=\left[
\begin{array}{ccc}
\mu_1 & 0 & 0 \\ 0 & \mu_2 & 0 \\ 0 & 0 & \mu_3
\end{array}
\right], \quad \mu_1\geq\mu_2\geq\mu_3\geq 0. \label{svdE}
\end{equation}
If the determinant of $G'$ is positive then one can choose $O_1$ and $O_2$
as proper orthogonal matrices (i.e.\ with the determinants equal to one).
In this case the singular value decomposition (\ref{svdE}) corresponds to a
local transformation $W=U_1\otimes U_2 W (U_1\otimes U_2)^\dagger$. In the
opposite case of a negative determinant of $G'$ one of the matrices
$O_1,O_2$
has also a negative determinant. Alternatively, we can assume that $O_1$ and
$O_2$ are proper orthogonal matrices (with positive determinants), and,
consequently, the singular value decomposition corresponds to a local
transformation, but with $\mu_1\leq\mu_2\leq\mu_3\leq 0$.

From (\ref{AD}) and (\ref{B}) it follows, that the above transformation
$G=O_1G'O_2^T$, if supplemented by ${\bf a}=O_1{\bf a'}$ and ${\bf
b}=O_2{\bf
b'}$, induces the transformation $C=C'(G',{\bf a'},{\bf b'})\mapsto C(G,{\bf
a},{\bf b})=(O_1\oplus O_2)C'(O_1\oplus O_2)^T$, where
\begin{equation}
O_1\oplus O_2:=\left[
\begin{array}{cc}
O_1 & 0 \\ 0 & O_2
\end{array}
\right],
\end{equation}
leaving the spectrum of $C$ invariant. The explicit form of the transformed
matrix inferred from (\ref{AD}), (\ref{B}), and (\ref{svdE}) reads
\begin{eqnarray}
C&=&\left[
\begin{array}{cccccc}
8\left(\mu_2^2+\mu_3^2\right) & 0 & 0 & \mp 16\mu_2\mu_3 & 0 & 0 \\ 0 &
8\left(\mu_1^2+\mu_2^3\right) & 0 & 0 & \mp 16\mu_1\mu_3 & 0 \\ 0 & 0 &
8\left(\mu_1^2+\mu_2^2\right) & 0 & 0 & \mp 16\mu_1\mu_2 \\ \mp 16\mu_2\mu_3
&
0 & 0 & 8\left(\mu_2^2+\mu_3^2\right) & 0 & 0 \\ 0 & \mp 16\mu_1\mu_3 & 0 &
0 &
8\left(\mu_1^2+\mu_3^2\right) & 0 \\ 0 & 0 & \mp 16\mu_1\mu_2 & 0 & 0 &
8\left(\mu_1^2+\mu_2^2\right)
\end{array}
\right] \nonumber \\ &+&\left[
\begin{array}{cc}
8\left(\|{\bf a}\|^2\cdot I-{\bf a}({\bf a})^T\right) & 0 \\ 0 &
8\left(\|{\bf
b}\|^2\cdot I-{\bf b}({\bf b})^T\right)
\end{array}
\right]:=C_G+C_{\bf{a},\bf{b}} \label{Cp}
\end{eqnarray}
which is the sum of two real positive definite matrices, $C_G$ and
$C_{\bf{a},\bf{b}}$. Their eigenvalues are, respectively
\begin{eqnarray}
\rho_1=8(\mu_1+\mu_2)^2,\; \rho_2=8(\mu_1+\mu_3)^2,\;
\rho_3=8(\mu_2+\mu_3)^2,
\nonumber \\ \rho_4=8(\mu_1-\mu_2)^2,\; \rho_5=8(\mu_1-\mu_3)^2,\;
\rho_6=8(\mu_2-\mu_3)^2, \label{rhos}
\end{eqnarray}
and
\begin{equation}
\nu_1=\nu_2=\|{\bf a}\|^2,\;  \nu_3=\nu_4=\|{\bf b}\|^2,\; \nu_5=\nu_6=0.
\label{nus}
\end{equation}

Although two parts, $C_G$ and $C_{\bf{a},\bf{b}}$ of $C$, usually, do not
commute and the eigenvalues $\lambda_1\geq\cdots\geq\lambda_6\geq 0$ of $C$
cannot be immediately found, we can investigate the possible orbits of
submaximal dimensionalities using the fact that
both $C_G$ and $C_{\bf{a},\bf{b}}$ are positive definite. It follows thus
that
the number of zero values among the eigenvalues $\lambda_1,\ldots,\lambda_6$
of
$C$ has to be matched by at least the same number of zeros among
$\rho_1,\ldots,\rho_6$ and among $\nu_1,\ldots,\nu_6$, moreover the
eigenvectors to the zero eigenvalues of the whole matrix $C$ are also the
eigenvectors of the components $C_G$ and $C_{\bf{a},\bf{b}}$ (also,
obvoiusly,
corresponding to the vanishing eigenvalues)

The co-rank ${r'}_{C_G}$ (the number of vanishing eigenvalues) of $C_G$
equals
\begin{eqnarray}
6 \quad &\text{for }\; & \mu_1=\mu_2=\mu_3=0 \Leftrightarrow G=0 , \nonumber
\\
3 \quad &\text{for }\; & \mu_1=\mu_2=\mu_3:=\mu\ne 0 \Leftrightarrow G=\mu
I,
\nonumber \\ 2 \quad &\text{for }\; & \mu:=\mu_1>\mu_2=\mu_3=0,
 \nonumber  \\
1 \quad &\text{for }\; & \mu_M:=\mu_1>\mu_2=\mu_3:=\mu_m\ne 0, \nonumber \\
        &\text{or  }\; & 0\ne\mu:=\mu_1=\mu_2>\mu_3,
        \label{E1b}
\end{eqnarray}
and is equal $0$ in all other cases, whereas for $C_{{\bf a},{\bf b}}$ it
co-rank ${r'}_{C_{{\bf a},{\bf b}}}$ reads
\begin{eqnarray}
6 \quad &\text{for }\; &{\bf a}={\bf b}=0, \nonumber \\ 4 \quad
&\text{for }\;
&{\bf a}=0,\;{\bf b}\ne 0 \; \text{or }\; {\bf a}\ne 0,\;{\bf b}=0,
\nonumber
\\ 2 \quad &\text{for }\; &{\bf a}\ne 0,\;{\bf b}\ne 0. \label{ab2}
\end{eqnarray}

As already mentioned, in a generic case all eigenvalues of the $6$D Gram
matrix
$C$ are positive and the dimension of local orbits is maximal, $d_l=6$. On
the
other hand, the above decomposition of the Gram matrix is very convenient to
analyze several special cases, for which some eigenvalues of $C$ reduce to
zero
and the local orbits are less dimensional. To find all of them  one needs to
consider $9$ combinations of different ranks of the matrices $C_G$ and
$C_{\bf{a},\bf{b}}$ as shown in the Appendix.

For any point of the Weyl chamber we know thus the dimension $D_g$ of the
corresponding global orbit. Using above results for any of the globally
equivalent states $W$ (with the same spectrum) we may find the dimension
$D_l$
of the corresponding local orbit. This dimension may be state dependent, as
explicitly shown for the case of $N=4$ pure states. Let $D_m$ denotes the
maximal dimension $D_l$, where the maximum is taken over all states of the
global orbit. The set of effectively different states, which {\sl cannot }
be
linked by local transformations has thus dimension $D_d=D_g - D_m$. For
example, the effectively different space of the $N=4$ pure states is one
dimensional, $D_d=6-5=1$.

\subsection{Special case: triple degeneracy and generalized Werner
states}

Consider the longest edge, $K_{13}$, of the Weyl chamber, which represents a
class of states with the triple degeneracy. They may be written in the form
$\rho_x:= x |\Psi\rangle\langle \Psi|+ (1-x)\rho_*$, where $|\Psi\rangle$
stands for {\sl any} pure state and $x\in (0,1]$. The global orbits have the
structure $U(4)/[U(3)\times U(1)]$, just as for the pure states, which are
generated by the corner of the simplex, represented by $x=1$. Also the
topology of the local orbits do not depends on $x$,
and the stratification found for pure states holds for each $6$ dimensional
global orbit generated by any single point of the edge.

Schematic drawing shown in  Fig.1 is still valid, but now the term
"maximally
entangled" denotes the entanglement maximal on the given global orbit. It
decreases with $x$ as for Werner states, with $|\Psi\rangle$ chosen as the
maximally entangled pure state  \cite{We89}. For these states the
concurrence
decreases linearly, $c(x)=(3x-1)/2$ for $x>1/3$ and is equal to zero for $x
\le
1/3$. Thus for sufficiently small $x$ (sufficiently large degree of mixing)
all
states are separable, also these belonging to one of the both $3$D local
orbits. This is consistent with the results of \cite{ZHSL}, where it was
proved
that if Tr$\rho^2<1/3$ the $2 \times 2$ mixed state $\rho$ is separable.

This condition has an appealing geometric interpretation: on one hand it
represents the maximal $3$D ball inscribed in the tetrahedron of
eigenvalues,
as shown in Fig.3. On the the other, it represents the maximal $15$D ball
$B_M$, (in sense of the Hilbert-Schmidt metric, $D_{HS}^2(\rho_1,\rho_2)=
\rm{Tr}(\rho_1-\rho_2)^2$), contained in the $15$D set of all mixed states
for
$N=4$. Both balls are centered at the maximally mixed state $\rho_*$ (the
center of the eigenvalues simplex of side $\sqrt{2}$), and have the same
radius
$1/2\sqrt{3}$. A similar geometric discussion of the properties of the set
of
$2 \times 2$ separable mixed states was recently given in \cite{DSZH00}.

\begin{figure}
  \hspace*{4.3cm}
 \epsfxsize=8.0cm
\epsfbox{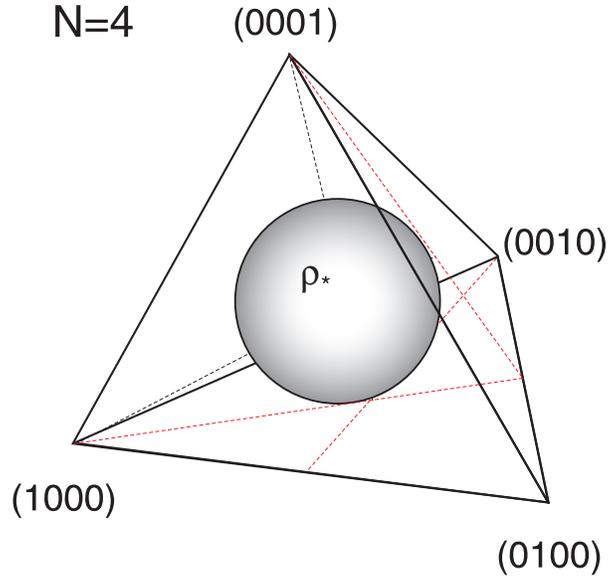} \\ \caption{ Separability of the maximal $15$D ball:
all
mixed states with spectra represented by points inside the ball inscribed in
the $3$D simplex of eigenvalues of the $N=4$ density matrices are
separable.}
\label{fig4}
\end{figure}

To clarify the structure of effectively different states in this case we
consider generalized Werner states
\begin{equation}
\rho(x,\theta):= x |\Psi_{\theta} \rangle\langle \Psi_{\theta}| +
(1-x)\rho_*,
\label{wern2}
\end{equation}
where the state
$|\Psi_{\theta}\rangle:=[\cos(\theta/2),0,0,\sin(\theta/2) ]$,
contains the line of effectively different pure states for $\theta \in
[0,\pi/2]$. Note that the case $\theta=\pi/2$ is equivalent to the original
Werner states \cite{We89}. Entanglement of formation  $E$ for the states
$\rho(x,\theta)$ may be computed analytically with help of concurrence and
the
Wootters formula \cite{Wo98}. The results are too lengthy to be reproduced
here, so in Fig.4 we present the plot $E=E(x,\theta)$. The graph is done in
polar coordinates, so the pure states are located at the circle $x=1$. For
each
fixed $x$ the space of effectively different states is represented by a
quarter
of the circle. For $x<1/3$ entire circle is located inside the maximal ball
$B_M$, and all effectively different states are separable.
Points located along a circle centerd at $\rho_*$ represent mixed states,
which are described by the same spectrum and can be connected
by a global unitary transformation $U(4)$.
In accordance to the recent results of Hiroshima and Ishizaka \cite{HI00},
the original Werner states enjoy the largest entanglement
accesible by unitary operations.

The convex set $\cal S$ of separable states contains a great section of the
maximal ball and touches the set of pure states in two points only. The
actual
shape of $\cal S$ (at this cross-section) looks remarkably similar to the
schematic drawing which appeared in \cite{Z99}. Moreover, the contour lines
of
constant $E$ elucidate important feature of any measure of entanglement: the
larger shortest distance to $\cal S$, the larger entanglement \cite{VP98}.
Even
though we are not going to prove that for any state $\rho$, its shortest
distance to $\cal S$ at the picture is strictly the shortest in the entire
$15$D space of mixed states, the geometric structure of the function
$E=E(x,\theta)$ is in some sense peculiar: The contours $E=$const are
foliated
along the boundary of $\cal S$, while  both maximally entangled
 states are located as far from $\cal S$, as possible.

\begin{figure}
  \hspace*{3.3cm}
 \epsfxsize=10.0cm
\epsfbox{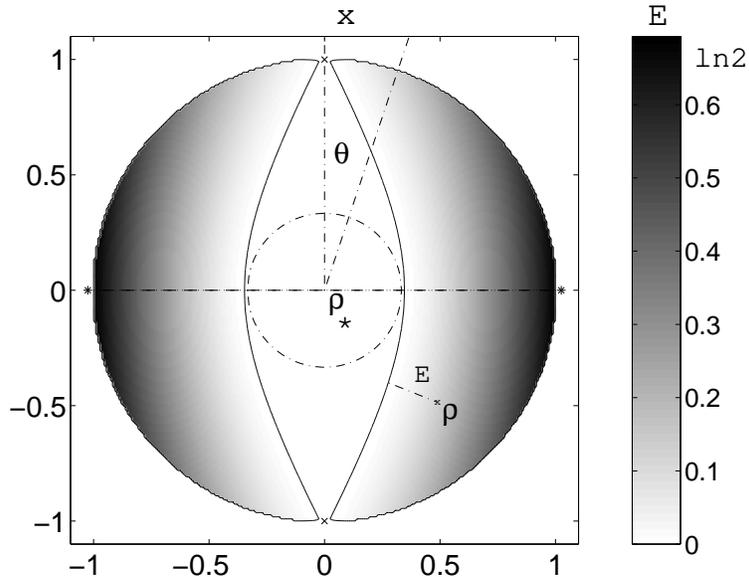} \\
 \caption{Entanglement of formation $E$ for the
generalized Werner states $\rho_{x,\theta}$ represented in the polar
coordinates. Intersection with the maximal ball centered at $\rho_*$ is
separable (white). Dashed horiznotal line, joining two maximally entangled
states $(*)$, (black), represents the original Werner states. Entanglement
$E$
of a mixed state $\rho$ may be interpreted as its distance from the set of
separable states. } \label{fig5}
\end{figure}

\subsection{ Absolutely separable states}

Defining {\sl separability} of a given mixed state $\rho$, we implicitly
assume
that the product structure of the composite Hilbert space is given, $\cal{
H}=\cal{ H}_A \otimes \cal{H}_B$. This assumption is well justified from the
physical point of view. For example, the EPR scenario distinguishes both
subsystems in a natural way ('left photon' and 'right photon'). Then we
speak
about separable (entangled) states, with respect to this particular
decomposition of $\cal H$. Note that any separable pure state may be
considered
entangled, if analyzed with respect to another decomposition of $\cal H$.

On the other hand, one may pose a complementary question, interesting merely
from the mathematical point of view, which states are separable with respect
to
{\sl any} possible decomposition of the $N=K \times M$ dimensional Hilbert
space $\cal H$. More formally, we propose the following

\smallskip
{\sl definition}. Mixed quantum state $\rho$ is called  {\sl absolutely
separable}, if all globally similar states $\rho'=U\rho U^{\dagger}$ are
separable.
\medskip

Unitary matrix $U$ of size $N$ represents a global operation equivalent to a
different choice of both subsystems. It is easy to see that the most mixed
state $\rho_*$ is absolutely separable. Moreover, the entire maximal ball
$B_M=B(\rho_*, 1/2\sqrt{3})$ is absolutely separable for $N=4$. This is
indeed
the case, since the proof of separability of  $B_M$ provided in \cite{ZHSL}
relays only on properties of the spectrum of $\rho$, invariant with respect
to global operations $U$. Another much simpler proof 
of separability of $B_M$ follows directly from inequality (9.21)
of the book of Mehta \cite{MM89}.

Are there any $2 \times 2 $ absolutely separable states not belonging to the
maximal ball $B_M$? Recent results of Ishizaka and Hiroshima \cite{IH00}
suggest, that this might be the case. They conjectured that the maximal
concurrence on the local orbit determined by the spectrum
$\{r_1,r_2,r_3,r_4\}$
is equal to $c^*={\rm max} \{0,r_1-r_3-2\sqrt{r_2r_4} \}$. This conjecture
has
been proved for the density matrices of rank $1,2$ and $3$ \cite{IH00}. If
it
is true in the general case than the condition $c^*>0$ defines the $3D$ set
of
spectra of absolutely separable states. This set belongs to the regular
tetrahedron of eigenvalues and contains the maximal ball $B_M$. For example,
a
state with the spectrum $\{0.47,0.30,0.13,0.10 \}$ does not belong to $B_M$
but
its $c^*$ is equal to zero.

\section{Concluding remarks }

In order to analyze geometric features of quantum entanglement we studied
the
properties of orbits generated by local transformations. Their shape and
dimensionality is not universal, but depends on the initial state. For each
quantum state of arbitrary $K \times M$ problem we defined the Gram matrix
$C$,
the spectrum of which remains invariant under local transformation. The rank
of
$C$ determines dimensionality of the local orbit. For generic mixed states
the
rank is maximal and equal to $D_l=K^2+M^2-2$,
while the space of all globally equivalent states (with the same spectrum)
is $(KM)^2-KM$ dimensional.
 Thus the set of states
effectively different, which cannot be related by any local transformation,
has
$D_d=(KM)^2-KM-(K^2+M^2-2)$ dimensions.

For the pure states of the simplest $2 \times 2$ problem we have shown that
the
set of effectively different states is one dimensional. This curve may be
parametrized by an angle emerging in the Schmidt decomposition: it  starts
at a
$3$D set of maximally entangled states, crosses the $5$D spaces of states of
gradually decreasing entanglement, and ends at the $4$D manifold of
separable states.

We presented an explicit parametrization of these submaximal manifolds.
Moreover, we have proved that any  pure state can be transformed by means of
local transformations into one of the states at this line. In such a way we
found a stratification of the $6$D manifold ${\mathbb C}P^3$ along the line
of
effectively different states into subspaces of different dimensionality.

Since for $N=4$ pure states the set of effectively different states is one
dimensional, all measures of entanglement must be equivalent (and be
functions
of, say, concurrence or entropy of formation). This is not the case for
generic
mixed states, for which $D_d=6$. Hence there exist mixed states of the same
entanglement of formation with the same spectrum (globally equivalent),
which
cannot be connected by means of local transformations.

It is known that some measures of entangled do not coincide (e.g.
entanglement
of formation $E$ and distillable entanglement $E_d$ \cite{Ho99}). To
characterize the entanglement of such mixed states one might, in principle,
use
$6$ suitably selected local invariants. This seem not to be very practical,
but
especially for higher systems, for which the dimension $D_d$, of effectively
different states is large and the bound entangled states exist (with $E_d=0$
and $E>0$), one may consider using some additional measures of entanglement.
All such measures of entanglement have to be functions of eigenvalues of the
Gram matrix $C$ or other invariants of local transformations
\cite{GRB98,LPS99,EM99,Su00,Ma00}.

We analyzed geometry of the convex set of separable states. For the simplest
$N=4$ problem it contains the maximal $15$D ball, inscribed in the set of
the
mixed states. It corresponds to the $3$D ball of radius $1/2\sqrt{3}$
inscribed
in the simplex of eigenvalues. This property holds also for $2\times 3$
problem, for which the radius is $1/\sqrt{30}$. For larger problems $K
\times
M=N\ge 8$, it is known that all mixed states in the maximal ball (of radius
$(N(N-1))^{-1/2}$) are not distillable \cite{ZHSL}, but the question whether
they are separable remains open.

{\bf Acknowledgments}

It is a pleasure to thank Pawe{\l}~Horodecki for several crucial comments
and Ingemar Bengtsson, Pawe{\l} Masiak and Wojciech S{\l}omczy{\'n}ski for
inspiring discussions.
 One of us (K.{\.Z}.) would like to thank the European Science
Foundation and the Newton Institute for a support allowing him to
participate in
the Workshop on {\sl Quantum Information} organized in Cambridge in July
1999,
where this work has been initiated. Financial support by a research grant 2
P03B
044 13 of Komitet Bada{\'n} Naukowych is gratefully acknowledged.

\appendix

\section{Submaximal local orbits for $2\times 2$ problem}

In this appendix we give the list of all possible submaximal ranks of the
Gram
matrix $C$ which determine the dimension of the local  orbit $D_l=6-r_C$.
The
symbol $r'_X$ denotes the co-rank, it is the number of zeros in the spectrum
of
$X$. In each submaximal case we provide the density matrix $W$, Gram matrix
$C$ and
its eigenvalues $\lambda_i$, $i=1,\dots,6$ expressed as a function of the
the singular values of the matrix  $G'$ and the vectors ${\bf a}=O_1 {\bf
a'}$
and ${\bf b}=O_2 {\bf b'}$, where orthogonal matrices $O_1$ and $O_2$ are
determined by the singular value decomposition of $G'$.

In a general case the density matrix $W=W\left(G ,{\bf a},{\bf b}\right)
=W\left( \mu _{1},\mu _{2},\mu
_{3},a_{1},a_{2},a_{3},b_{1},b_{2},b_{3}\right)
$ is given by
\begin{equation}
W=\ \ \frac{1}{4}I+\left[
\begin{array}{cccc}
-a_{3}-b_{3}-\mu _{3} & -b_{1}-ib_{2} & -a_{1}-ia_{2} & -\mu _{1}+\mu _{2}
\\
-b_{1}+ib_{2} & -a_{3}+b_{3}+\mu _{3} & -\mu _{1}-\mu _{2} & -a_{1}-ia_{2}
\\
-a_{1}+ia_{2} & -\mu _{1}-\mu _{2} & a_{3}-b_{3}+\mu _{3} & -b_{1}-ib_{2} \\
-\mu _{1}+\mu _{2} & -a_{1}+ia_{2} & -b_{1}+ib_{2} & a_{3}+b_{3}-\mu _{3}
\end{array}
\right] \label{ap1}
\end{equation}
where we use the rotated basis in which $G$ is diagonal. 
The characteristic equation of the density matrix $W$ reads
\begin{eqnarray}
\det\left(W-\varrho\right)&=&\varrho^4-\varrho^3+
\left[\frac{3}{8}-2\left\|{\bf a}\right\|^2-2\left\|{\bf b}\right\|^2
-2\text{Tr}G^2\right]\varrho^2+\left(-\frac{1}{16}+
\left\|{\bf a}\right\|^2+\left\|{\bf b}\right\|^2
+\text{Tr}G^2+8{\bf a}G{\bf b}-8\det G\right)\varrho
\nonumber \\
&&+\left(\left\|{\bf a}\right\|^2-\left\|{\bf b}\right\|^2\right)^2
+2\text{Tr}G^4-\left(\text{Tr}G^2\right)^2
-\frac{1}{8}\left\|{\bf a}\right\|^2-\frac{1}{8}\left\|{\bf b}\right\|^2
-\frac{1}{8}\text{Tr}G^{2}-2 {\bf a}G{\bf b}+2\det G
\nonumber \\
&&
-4\left\|G{\bf a}\right\|^2-4\left\|G{\bf b}\right\|^2
+2\left(\left\|{\bf a}\right\|^2+\left\|{\bf b}\right\|^2\right)\text{Tr}G^2
+8\left(a_1b_1\mu_2\mu_3+a_2b_2\mu_1\mu_3+a_3b_3\mu_1\mu_2\right)
+\frac{1}{256}.
\label{char1}
\end{eqnarray}
It is interesting to note that the characteristic equation of the partially
transposed matrix $\widetilde{W}=W^{T_2}$ differs only by signs of three terms:
\begin{eqnarray}
\det\left(W-\widetilde{\varrho}\right)&=&\widetilde{\varrho}^4-
\widetilde{\varrho}^3+\left[\frac{3}{8}-2\left\|{\bf a}\right\|^2
-2\left\|{\bf b}\right\|^2-2\text{Tr}G^2\right]\widetilde{\varrho}^{2}
+\left(-\frac{1}{16}+\left\|{\bf a}\right\|^2+\left\|{\bf b}\right\|^2
+\text{Tr}G^{2}+8{\bf a}G{\bf b}+8\det G\right)
\widetilde{\varrho}
\nonumber \\
&&+\left(\left\|{\bf a}\right\|^{2}-\left\|{\bf b}\right\|^{2}\right)^{2}
+2\text{Tr}G^{4}-\left( \text{Tr}G^{2}\right)^{2}
-\frac{1}{8}\left\|{\bf a}\right\|^2-\frac{1}{8}\left\|{\bf b}\right\|^2
-\frac{1}{8}\text{Tr}G^{2}-2{\bf a}G{\bf b}-2\det G
\nonumber \\
&&-4\left\| G{\bf a}\right\|^{2}-4\left\| G{\bf b}\right\|^{2}
+2\left(\left\|{\bf a}\right\|^2+\left\|{\bf b}\right\|^2\right)\text{Tr}G^{2}
-8\left(a_1b_1\mu_2\mu_3+a_2b_2\mu_1\mu_3+a_3b_3\mu_1\mu_{2}\right)
+\frac{1}{256}.
\label{char2}
\end{eqnarray} 
Let $\varrho _{i}$ and $\widetilde{\varrho }_{i}$, $i=1,2,3,4$, denote the
eigenvalues of $W$ and  $\widetilde{W}$, respectively. Due to
Peres-Horodeccy
partial transpose criterion \cite{Pe96,Ho96} positivity of
$\widetilde{\varrho
}_{i}$ may be used to find, under which conditions $W$ is separable.

In order to compute the concurrence of the density matrix $W$, let us define
an
auxiliary hermitian matrix
\begin{equation}
\overline{W}:=W\sigma _{2}\otimes \sigma _{2}W^{\ast }\sigma _{2}\otimes
\sigma
_{2}, \label{wbar}
\end{equation}
where $^*$ represents the complex conjugation. Let $\xi _{i}$, $i=1,2,3,4$
denote the eigenvalues of $\overline{W}$, arranged in decreasing order. Then
the concurrence $c$ of $W$ is given by \cite{HW97,Wo98}
\begin{equation}
c:=\max \left( 0,\sqrt{\xi _{1}}-\sqrt{\xi _{2}}-\sqrt{\xi _{3}}-\sqrt{\xi
_{4}}\right). \label{conc3}
\end{equation}

The Gram matrix $C=C\left( G,{\bf a},{\bf b}\right) =C\left( \mu _{1},\mu
_{2},\mu _{3},a_{1},a_{2},a_{3},b_{1},b_{2},b_{3}\right)$ corresponding to
the
density matrix $W$, reads in the general case
\begin{equation}
C=8\left[
\begin{array}{cccccc}
a_{2}^{2}+a_{3}^{2}+\mu _{2}^{2}+\mu _{3}^{2} & -a_{1}a_{2} & -a_{1}a_{3} &
-2\mu _{2}\mu _{3} & 0 & 0 \\ -a_{1}a_{2} & a_{1}^{2}+a_{3}^{2}+\mu
_{1}^{2}+\mu _{3}^{2} & -a_{2}a_{3} & 0 & -2\mu _{1}\mu _{3} & 0
\\ -a_{1}a_{3}
& -a_{2}a_{3} & a_{1}^{2}+a_{2}^{2}+\mu _{1}^{2}+\mu _{2}^{2} & 0 & 0
& -2\mu
_{1}\mu _{2} \\ -2\mu _{2}\mu _{3} & 0 & 0 & b_{2}^{2}+b_{3}^{2}+\mu
_{2}^{2}+\mu _{3}^{2} & -b_{1}b_{2} & -b_{1}b_{3} \\ 0 & -2\mu _{1}\mu _{3}
& 0
& -b_{1}b_{2} & b_{1}^{2}+b_{3}^{2}+\mu _{1}^{2}+\mu _{3}^{2} & -b_{2}b_{3}
\\
0 & 0 & -2\mu _{1}\mu _{2} & -b_{1}b_{3} & -b_{2}b_{3} &
b_{1}^{2}+b_{2}^{2}+\mu _{1}^{2}+\mu _{2}^{2}
\end{array}
\right] \label{ccc}
\end{equation}

Below we provide a list of the classes of states corresponding to the
submaximal ranks $r_C$  of the Gram matrices. The list is ordered according
to
the increasing dimensionality of local orbits; $D_l=r_C=6-r_{C}^{\prime}$.

\medskip
{\bf Case 1.} $r_{C}^{\prime }=6,$ \ $G=0,$ \ ${\bf a}=0,$ \ ${\bf b}=0,$;
$C=0$
\begin{equation}
\lambda _{1,2,3,4,5,6}=0; ~ ~ W=\frac{1}{4}I; ~ ~ \varrho
_{1,2,3,4}=\frac{1}{4}=\widetilde{\varrho }_{1,2,3,4}, ~ ~ \xi
_{1,2,3,4}=\frac{1}{16}, \label{c1l}
\end{equation}
thus $W$ is separable and concurrence, $c$, is equal to zero.

\medskip
{\bf Case 2.}  $r_{C}^{\prime }=4,$ \ $G=0,$ \ ${\bf a}\ne 0,$ \ ${\bf
b}=0,$
\begin{equation}
\lambda _{1,2}=8\Vert {\bf a}\Vert ^{2},~ ~ \lambda _{3,4,5,6}=0;
\label{c2la}
\end{equation}
\begin{equation}
\varrho _{1,2}=\frac{1}{4}+\left\| {\bf a}\right\| ,\varrho
_{3,4}=\frac{1}{4}-\left\| {\bf a}\right\| ; \label{c2r1}
\end{equation}
\begin{equation}
\widetilde{\varrho }_{1,2}=\frac{1}{4}+\left\| {\bf a}\right\| ,\widetilde{%
\varrho }_{3,4}=\frac{1}{4}-\left\| {\bf a}\right\| . \label{c2r2}
\end{equation}
\begin{equation}
\xi _{1,2,3,4}=\frac{1}{16}-\left\| {\bf a}\right\| ^{2}, ~ ~ {\rm thus} ~ ~
c=0. \label{c2c2}
\end{equation}
$W$ represents a density matrix for $\left\| {\bf a}\right\| \leq
\frac{1}{4}$
and then is separable ($\widetilde{W} \ge 0 $).

\medskip
{\bf Case 3.}  $r_{C}^{\prime }=3$; $G=\mu I,$ \ ${\bf a}=0,$ \ ${\bf b}=0,$
\begin{equation}
\lambda_{1,2,3}=32\mu ^{2},\ \lambda _{4,5,6}=0 \label{c3l}
\end{equation}
\begin{equation}
\varrho _{1,2,3}=\frac{1}{4}-\mu ,\ \varrho _{4}=\frac{1}{4}+3\mu,
\label{c3r1}
\end{equation}
\begin{equation}
\widetilde{\varrho }_{1,2,3}=\frac{1}{4}+\mu ,\ \widetilde{\varrho }_{4}=%
\frac{1}{4}-3\mu. \label{c3r2}
\end{equation}
\begin{equation}
\xi _{1}=\frac{1}{16}\left( 12\mu +1\right) ^{2},\quad \xi
_{2,3,4}=\frac{1}{%
16}\left( 1-4\mu \right) ^{2}, \label{c3c2}
\end{equation}
\begin{equation}
c=\left\{
\begin{array}{ccc}
0 & \text{for} & \mu \leq \frac{1}{12} \\ 6\mu -\frac{1}{2} &
\text{for}
& \frac{1}{12}\leq \mu \leq \frac{1}{4}
\end{array}
\right. \label{c3c3}
\end{equation}
$W\geq 0$ \ for \ $-\frac{1}{12}\leq\mu \leq \frac{1}{4}$ and
$W$ is separable for $|\mu|\leq \frac{1}{12}$.

\medskip
{\bf Case 4.}  $r_{C}^{\prime }=2$.
$G=0,$ \
\begin{equation}
\lambda _{1,2}=8\Vert {\bf a}\Vert ^{2},\ \lambda _{3,4}=8\Vert {\bf b}\Vert
^{2},\lambda _{5,6}=0. \label{c4l}
\end{equation}
\begin{equation}
\varrho _{1}=\frac{1}{4}+\left\| {\bf a}\right\| +\left\| {\bf b}\right\| ,\
\varrho _{2}=\frac{1}{4}-\left\| {\bf a}\right\| -\left\| {\bf b}\right\| ,\
\ \varrho _{3}=\frac{1}{4}+\left| \left\| {\bf a}\right\| -\left\| {\bf b}%
\right\| \right| ,\ \ \varrho _{4}=\frac{1}{4}-\left| \left\| {\bf a}%
\right\| -\left\| {\bf b}\right\| \right| \label{c4r1}
\end{equation}
\begin{equation}
\widetilde{\varrho }_{1}=\frac{1}{4}+\left\| {\bf a}\right\| +\left\| {\bf
b}%
\right\| ,\ \widetilde{\varrho }_{2}=\frac{1}{4}-\left\| {\bf a}\right\|
-\left\| {\bf b}\right\| ,\ \ \widetilde{\varrho }_{3}=\frac{1}{4}+\left|
\left\| {\bf a}\right\| -\left\| {\bf b}\right\| \right| ,\ \ \widetilde{%
\varrho }_{4}=\frac{1}{4}-\left| \left\| {\bf a}\right\| -\left\| {\bf b}%
\right\| \right| . \label{c4r2}
\end{equation}
\begin{equation}
\xi _{1,2}=\frac{1}{16}+\left( \left\| {\bf a}\right\| +\left\| {\bf b}%
\right\| \right) ^{2},\quad \xi _{3,4}=\frac{1}{16}+\left( \left\| {\bf a}%
\right\| -\left\| {\bf b}\right\| \right) ^{2}; \quad c=0. \label{c4c2}
\end{equation}

$W\geq 0$ \ for \ $\left\| {\bf a}\right\| +\left\| {\bf b}\right\| \leq
\frac{1}{4}$ and is then separable.

\medskip
{\bf Case 5.}  ${r^{\prime }}_{C}=2$

$G=diag(\mu ,0,0),$ \ ${\bf a}=\left[ a,0,0\right] ^{T},$ \ ${\bf b}=\left[
b,0,0\right] ^{T},$
\begin{equation}
\lambda _{1,2}=8\left( a^{2}+\mu ^{2}\right) ,\quad \lambda _{3,4}=8\left(
b^{2}+\mu ^{2}\right) ,\ \ \lambda _{5,6}=0 \label{c5l}
\end{equation}
\begin{equation}
\varrho _{1}=\frac{1}{4}+a+b-\mu ,\ \varrho _{2}=\frac{1}{4}-a+b+\mu ,\
\varrho
_{3}=\frac{1}{4}-a-b-\mu ,\ \varrho _{4}=\frac{1}{4}+a-b+\mu , \label{c5r1}
\end{equation}
\begin{equation}
\widetilde{\varrho }_{1}=\frac{1}{4}+a+b-\mu ,\ \widetilde{\varrho }_{2}=%
\frac{1}{4}-a+b+\mu ,\ \widetilde{\varrho }_{3}=\frac{1}{4}-a-b-\mu ,\
\widetilde{\varrho }_{4}=\frac{1}{4}+a-b+\mu . \label{c5r2}
\end{equation}
\begin{equation}
\xi _{1,2}=\left( \frac{1}{4}+\mu \right) ^{2}-\left( a-b\right) ^{2}, \quad
\xi _{3,4}=\left( \frac{1}{4}-\mu \right) ^{2}-\left( a+b\right) ^{2}, \quad
c=0. \label{c5c2}
\end{equation}

$W\geq 0$ \ for \ $a=\left\| {\bf a}\right\| \leq \frac{1}{4}$, $\ b=\left\|
{\bf b}\right\| \leq \frac{1}{4}$, \ $|\mu| \leq \frac{1}{4}$; then $W$ is
separable.
\medskip

{\bf Case 6.} ${r^{\prime }}_{C}=1$

$G=diag(\mu ,0,0),$ \ ${\bf a}=\left[ a,0,0\right] ^{T},$ \
\begin{eqnarray}
\lambda _{1} &=&4\left( \Vert {\bf b}\Vert ^{2}+\mu ^{2}+\sqrt{(\mu
^{2}-\Vert
{\bf b}\Vert ^{2})^{2}+4\mu ^{2}{b}_{1}^{2}}\right) \nonumber \\ \lambda
_{2}
&=&4\left( \Vert {\bf b}\Vert ^{2}+\mu ^{2}-\sqrt{(\mu ^{2}-\Vert {\bf
b}\Vert
^{2})^{2}+4\mu ^{2}{b}_{1}^{2}}\right) \nonumber \\ \lambda _{3} &=&8\left(
\Vert {\bf b}\Vert ^{2}+\mu ^{2}\right) ,\lambda _{4,5}=8\left( a^{2}+\mu
^{2}\right) ,\quad \lambda _{6}=0\ \ \label{c6l}
\end{eqnarray}
\begin{eqnarray}
\varrho _{1} &=&\frac{1}{4}+a+\sqrt{\mu ^{2}+\left\| {\bf b}\right\|
^{2}-2b_{1}\mu },\ \ \varrho _{2}=\frac{1}{4}+a-\sqrt{\mu ^{2}+\left\| {\bf
b}\right\| ^{2}-2b_{1}\mu }, \nonumber \\ \varrho _{3}
&=&\frac{1}{4}-a+\sqrt{\mu ^{2}+\left\| {\bf b}\right\| ^{2}+2b_{1}\mu },\ \
\varrho _{4}=\frac{1}{4}-a-\sqrt{\mu ^{2}+\left\| {\bf b}\right\|
^{2}+2b_{1}\mu }, \label{c6r1}
\end{eqnarray}
\begin{eqnarray}
\widetilde{\varrho }_{1} &=&\frac{1}{4}+a+\sqrt{\mu ^{2}+\left\| {\bf b}%
\right\| ^{2}-2b_{1}\mu },\ \ \widetilde{\varrho }_{2}=\frac{1}{4}+a-\sqrt{%
\mu ^{2}+\left\| {\bf b}\right\| ^{2}-2b_{1}\mu }, \nonumber \\
\widetilde{\varrho }_{3} &=&\frac{1}{4}-a+\sqrt{\mu ^{2}+\left\| {\bf b}%
\right\| ^{2}+2b_{1}\mu },\ \ \widetilde{\varrho }_{4}=\frac{1}{4}-a-\sqrt{%
\mu ^{2}+\left\| {\bf b}\right\| ^{2}+2b_{1}\mu }, \label{c6r2}
\end{eqnarray}
\begin{eqnarray}
\xi _{1,2} &=&\frac{1}{16}+\mu ^{2}-a^{2}-\left\| {\bf b}\right\|
^{2}+\sqrt{%
4\left( a^{2}-\mu ^{2}\right) \left\| {\bf b}\right\| ^{2}+4\mu
^{2}b_{1}^{2}+2\mu ab_{1}}, \nonumber \\
\xi _{3,4} &=&\frac{1}{16}+\mu ^{2}-a^{2}-\left\| {\bf b}\right\|
^{2}-\sqrt{%
4\left( a^{2}-\mu ^{2}\right) \left\| {\bf b}\right\| ^{2}+4\mu
^{2}b_{1}^{2}+2\mu ab_{1}},
\end{eqnarray}
so $c=0$. If $W$ represents a density matrix $(W\geq 0)$ then it is
separable.

\medskip
{\bf Case 7.} $r_{C}^{\prime }=1$

$G=\mu I,$ \ ${\bf b}=\xi {\bf a},$
\begin{eqnarray}
\lambda _{1} &=&4\left( \left( \xi ^{2}-1\right) \Vert {\bf a}\Vert ^{2}+%
\sqrt{\mu ^{2}+\sqrt{16\mu ^{4}+\left( \xi ^{2}-1\right) ^{2}\Vert {\bf a}%
\Vert ^{2}}}\right) , \nonumber \\
\lambda _{2} &=&4\left( \left( \xi ^{2}-1\right) \Vert {\bf a}\Vert ^{2}+%
\sqrt{\mu ^{2}-\sqrt{16\mu ^{4}+\left( \xi ^{2}-1\right) ^{2}\Vert {\bf a}%
\Vert ^{2}}}\right) , \nonumber \\
\lambda _{3} &=&4\left( \left( \xi ^{2}-1\right) \Vert {\bf a}\Vert ^{2}-%
\sqrt{\mu ^{2}+\sqrt{16\mu ^{4}+\left( \xi ^{2}-1\right) ^{2}\Vert {\bf a}%
\Vert ^{2}}}\right) , \nonumber \\
\lambda _{4} &=&4\left( \left( \xi ^{2}-1\right) \Vert {\bf a}\Vert ^{2}-%
\sqrt{\mu ^{2}-\sqrt{16\mu ^{4}+\left( \xi ^{2}-1\right) ^{2}\Vert {\bf a}%
\Vert ^{2}}}\right) , \nonumber \\ \lambda _{5} &=&32\mu ^{2},\lambda _{6}=0
\label{c7l}
\end{eqnarray}
\begin{eqnarray}
\varrho _{1} &=&\frac{1}{4}-\mu +\left| \xi +1\right| \left\| {\bf a}%
\right\| ,\ \ \varrho _{2}=\frac{1}{4}-\mu -\left| \xi +1\right| \left\|
{\bf
a}\right\| , \nonumber \\ \varrho _{3} &=&\frac{1}{4}+\mu +\sqrt{4\mu
^{2}+\left( \xi -1\right) ^{2}\left\| {\bf a}\right\| ^{2}},\ \varrho
_{4}=\frac{1}{4}+\mu -\sqrt{4\mu ^{2}+\left( \xi -1\right) ^{2}\left\| {\bf
a}\right\| ^{2}} \label{c7r1}
\end{eqnarray}
\begin{eqnarray}
\widetilde{\varrho }_{1} &=&\frac{1}{4}+\mu +\left| \xi -1\right| \left\|
{\bf
a}\right\| ,\ \ \widetilde{\varrho }_{2}=\frac{1}{4}+\mu -\left|
\xi -1\right|
\left\| {\bf a}\right\| , \nonumber \\ \widetilde{\varrho }_{3}
&=&\frac{1}{4}-\mu +\sqrt{4\mu ^{2}+\left( \xi +1\right) ^{2}\left\| {\bf
a}\right\| ^{2}},\
\widetilde{\varrho }_{4}=\frac{%
1}{4}-\mu -\sqrt{4\mu ^{2}+\left( \xi +1\right) ^{2}\left\| {\bf a}\right\|
^{2}} \label{c7r2}
\end{eqnarray}
\begin{eqnarray}
\xi _{1} &=&\frac{1}{16}+\frac{\mu }{2}+5\mu ^{2}-\left( \xi -1\right)
^{2}\left\| {\bf a}\right\| ^{2}+\mu \sqrt{4\left( \mu +1\right)
^{2}-16\left(
\xi -1\right) ^{2}\left\| {\bf a}\right\| ^{2}}, \nonumber \\ \xi _{2}
&=&\frac{1}{16}+\frac{\mu }{2}+5\mu ^{2}-\left( \xi -1\right) ^{2}\left\|
{\bf
a}\right\| ^{2}-\mu \sqrt{4\left( \mu +1\right) ^{2}-16\left( \xi -1\right)
^{2}\left\| {\bf a}\right\| ^{2}}, \nonumber \\ \xi _{3,4} &=&\left(
\frac{1}{4}-\mu \right) ^{2}-(\left( \xi +1\right) ^{2}\left\| {\bf
a}\right\|
^{2}
\end{eqnarray}
If $W\geq 0$ \ i.e. $\varrho _{i}\geq 0$, $i=1,2,3,4$ then $|\mu| \leq
\frac{1}{4}$ and $\widetilde{\varrho }_{i}\geq 0$, $i=1,3$, hence $W$ is
nonseparable for $\ \sqrt{4\mu ^{2}+\left( \xi +1\right) ^{2}\left\| {\bf
a}\right\| ^{2}}>\frac{1}{4}-\mu \geq \left| \xi +1\right| \left\| {\bf a}
\right\| $ or $\frac{1}{4}<|\xi-1|\|{\bf a}\|$.

\medskip
{\bf Case 8.}  $r_{C}^{\prime }=1$

$G=diag(\mu _{1},\mu _{2},\mu _{2}),$ \ ${\bf a}=\left[ a,0,0\right] ^{T},$
\
${\bf b}=\left[ b,0,0\right] ^{T},$
\begin{eqnarray}
\lambda _{1,2} &=&4\left( a^{2}+b^{2}+2\mu _{1}^{2}+2\mu _{2}^{2}+\sqrt{%
16\mu _{1}^{2}\mu _{2}^{2}+\left( a^{2}-b^{2}\right) ^{2}}\right) ,
\nonumber
\\
\lambda _{3,4} &=&4\left( a^{2}+b^{2}+2\mu _{M}^{2}+2\mu _{m}^{2}-\sqrt{%
16\mu _{1}^{2}\mu _{2}^{2}+\left( a^{2}-b^{2}\right) ^{2}}\right) ,\lambda
_{5}=32\mu _{2}^{2},\lambda _{6}=0
\end{eqnarray}
\label{c8l}
\begin{eqnarray}
\varrho _{1} &=&\frac{1}{4}-\mu _{1}+a+b,\ \ \varrho _{2}=\frac{1}{4}-\mu
_{1}-a-b, \nonumber \\ \varrho _{3} &=&\frac{1}{4}+\mu _{1}+\sqrt{4\mu
_{2}^{2}+\left( a-b\right) ^{2}},\ \ \varrho _{4}=\frac{1}{4}+\mu
_{1}-\sqrt{4\mu _{2}^{2}+\left( a-b\right) ^{2}}
\end{eqnarray}
\label{c8r1}
\begin{eqnarray}
\widetilde{\varrho }_{1} &=&\frac{1}{4}+\mu _{1}-a+b,\ \
\widetilde{\varrho }%
_{2}=\frac{1}{4}+\mu _{1}+a-b, \nonumber \\ \widetilde{\varrho }_{3}
&=&\frac{1}{4}-\mu _{1}+\sqrt{4\mu _{2}^{2}+\left(
a+b\right) ^{2}},\ \ \widetilde{\varrho }_{4}=\frac{1}{4}-\mu _{1}-\sqrt{%
4\mu _{2}^{2}+\left( a+b\right) ^{2}} \label{c8r2}
\end{eqnarray}

\begin{eqnarray}
\xi _{1,2} &=&\left( \frac{1}{4}-\mu _{1}\right)^{2} -\left( a+b\right)
^{2},
\quad \xi _{3}  = \left( \sqrt{\left( \frac{1}{4}+\mu _{1}\right)
^{2}-\left(
a-b\right) ^{2}}+2\mu _{2}\right) ^{2}, \nonumber  \\ \xi _{4} &=&\left(
\sqrt{\left( \frac{1}{4}+\mu _{1}\right) ^{2}-\left( a-b\right) ^{2}}-2\mu
_{2}\right) ^{2},
\end{eqnarray}
If $W\geq 0$ \ i.e. $\varrho _{i}\geq 0$, $i=1,2,3,4$ then $|\mu _{1}|\leq
\frac{1}{4}$ and $\widetilde{\varrho }_{i}\geq 0$, $i=1,3$, hence $W$ is
nonseparable for $\ \sqrt{4\mu _{2}^{2}+\left( a+b\right)^{2}}
> \frac{1}{4} -\mu _{1}\geq \left| a+b\right| $ or $\frac{1}{4}<b-a-\mu_1$.

\medskip
{\bf Case 9.}  $r_{C}^{\prime }=1$

$G=diag(\mu _{1},\mu _{1},\mu _{2}),$ \ ${\bf a}= \left[ 0,0,a\right] ^{T},$
\
${\bf b}=\left[ 0,0,b\right] ^{T}.$
\begin{eqnarray}
\lambda _{1,2} &=&4\left( a^{2}+b^{2}+2\mu _{1}^{2}+2\mu _{2}^{2}+\sqrt{%
16\mu _{1}^{2}\mu _{2}^{2}+\left( a^{2}-b^{2}\right) ^{2}}\right) ,
\nonumber
\\
\lambda _{3,4} &=&4\left( a^{2}+b^{2}+2\mu _{1}^{2}+2\mu _{2}^{2}-\sqrt{%
16\mu _{1}^{2}\mu _{2}^{2}+\left( a^{2}-b^{2}\right) ^{2}}\right) ,\lambda
_{5}=32\mu _{1}^{2},\lambda _{6}=0, \label{c9l}
\end{eqnarray}

\begin{eqnarray}
\varrho _{1} &=&\frac{1}{4}-\mu _{2}+a+b,\ \ \varrho _{2}=\frac{1}{4}-\mu
_{2}-a-b, \nonumber \\ \varrho _{3} &=&\frac{1}{4}+\mu _{2}+\sqrt{4\mu
_{1}^{2}+\left( a-b\right) ^{2}},\ \ \varrho _{4}=\frac{1}{4}+\mu
_{2}-\sqrt{4\mu _{1}^{2}+\left( a-b\right) ^{2}} \label{c9r1}
\end{eqnarray}
\begin{eqnarray}
\widetilde{\varrho }_{1} &=&\frac{1}{4}+\mu _{2}+a-b,\ \ \widetilde{\varrho
}_{2}=\frac{1}{4}+\mu _{2}-a+b, \nonumber \\ \widetilde{\varrho }_{3}
&=&\frac{1}{4}-\mu _{2}+\sqrt{4\mu _{1}^{2}+\left(
a+b\right) ^{2}},\ \ \widetilde{\varrho }_{4}=\frac{1}{4}-\mu _{2}-\sqrt{%
4\mu _{1}^{2}+\left( a+b\right) ^{2}} \label{c9r2}
\end{eqnarray}
\begin{eqnarray}
\xi _{1,2} &=&\left( \frac{1}{4}-\mu _{2}\right)^{2} -\left( a+b\right)^{2},
\quad \xi _{3} =\left( \sqrt{\left( \frac{1}{4}+\mu _{2}\right) ^{2}-\left(
a-b\right) ^{2}}+2\mu _{1}\right) ^{2}, \nonumber \\ \xi _{4} &=&\left(
\sqrt{\left( \frac{1}{4}+\mu _{2}\right) ^{2}-\left( a-b\right) ^{2}}-2\mu
_{1}\right) ^{2},
\end{eqnarray}
If $W\geq 0$ \ i.e. $\varrho _{i}\geq 0$, $i=1,2,3,4$ then $|\mu _{2}|\leq
\frac{1}{4}$ and $\widetilde{\varrho }_{i}\geq 0$, $i=1,3$, hence $W$ is
nonseparable for $\ \sqrt{4\mu _{1}^{2}+ \left( a+b\right)^{2}}>\frac{1}{4}
 -\mu _{2}\geq \left| a+b\right| $ or $\frac{1}{4}< b-a-\mu_2$.

Note that the the dimensionality $D_l$ given for each item holds for a
non-zero
choice of the relevant parameters. Some eigenvalues $\lambda_i$ may vanish
under a special choice of parameters - these subcases are easy to find.
There
exists also symmetric cases $2'$ and $6'$ for which the vectors $\bf a$ and
$\bf b$ are exchanged. The dimensionality of the local orbits remains
unchanged, and the formulae for eigenvalues hold, if one exchanges both
vectors.

\end{document}